\documentclass[letterpaper]{article}
\usepackage[preprint]{aaai2027}
\usepackage[hyphens]{url}
\usepackage{graphicx}
\usepackage{natbib}
\usepackage{caption}
\usepackage{booktabs}
\usepackage{amsfonts}
\usepackage{amsmath}
\usepackage{amssymb}
\usepackage{nicefrac}
\usepackage{microtype}
\usepackage{multirow}
\usepackage{array}
\usepackage{xcolor}
\usepackage{tikz}
\usepackage{pgfplots}
\pgfplotsset{compat=1.18}
\usepackage{placeins}

\newcommand{\UnpairGROverviewFigurePath}{figures/umagr.pdf}
\newcommand{\UnpairGRVenue}[1]{#1}
\newcommand{\UnpairGRGain}[1]{\textbf{#1}}
\newcommand{\UnpairGRBest}[1]{\textbf{#1}}
\newcommand{\UnpairGRSecond}[1]{\underline{#1}}

\title{Unpaired Modality-Agnostic Generative Recommendation}

\author{
    Weihao Shen\textsuperscript{\rm 1},
    Wei Chen\textsuperscript{\rm 1},
    Fuwei Zhang\textsuperscript{\rm 1},
    Meng Yuan\textsuperscript{\rm 1},
    Yuqin Lan\textsuperscript{\rm 1},\\
    Guojun Liu\textsuperscript{\rm 2},
    Qingsong Hua\textsuperscript{\rm 2},
    Wei Lin\textsuperscript{\rm 2},
    Fuzhen Zhuang\textsuperscript{\rm 1}\corresponding
}
\affiliations{
    \textsuperscript{\rm 1}Institute of Artificial Intelligence, Beihang University, Beijing, China\\
    \textsuperscript{\rm 2}Meituan, Beijing, China
}

\begin{document}

\maketitle

\begin{abstract}
Generative Recommendation (GR) formulates recommendation as autoregressive generation over discrete semantic identifiers (IDs). Although recent multimodal GR methods improve semantic ID construction with visual and textual information, they typically require item-level paired observations, restricting tokenization to the intersection of modality availability. Moreover, incorporating unpaired observations is nontrivial because small representation shifts may cross quantization boundaries and produce incompatible identifier sequences. To address this challenge, we propose \textbf{Unpair}ed Modality-Agnostic \textbf{G}enerative \textbf{R}ecommendation (UnpairGR), which learns a unified semantic-ID space from paired, image-only, and text-only observations. UnpairGR confines modality-specific processing to lightweight input projections while sharing the subsequent Transformer and residual codebooks across all observation conditions. Paired observations establish a reliability-guided cross-modal consensus, whereas unimodal observations directly refine the same representations and codes. The learned tokenizer is then fixed to provide stationary targets for a single autoregressive recommender, without feature imputation, modality-specific codebooks, or fallback mappings. Extensive experiments on three benchmark datasets demonstrate that UnpairGR consistently improves recommendation performance under both fully observed and incomplete-observation settings. 
\end{abstract}
\begin{links}
\link{Code}{https://github.com/Nevaeh7/UnpairGR}
\end{links}
\section{Introduction}
Generative Recommendation (GR) formulates next-item recommendation as autoregressive generation over discrete semantic identifiers (IDs)~\cite{rajput2023tiger,hou2025rpg,li2026lsig}. In this paradigm, each item is represented by a sequence of semantic tokens, and the recommender predicts the next item by generating its semantic ID from the user's interaction history. The effectiveness of GR therefore largely depends on whether these IDs can faithfully encode item semantics. If the IDs are derived from limited signals, semantically related items may be assigned unrelated token sequences, making it difficult for the generative model to learn meaningful item relationships~\cite{singh2024semanticids}.

To improve the quality of semantic IDs, recent multimodal GR methods~\cite{zhang2026macrec,chen2026syngr} incorporate multimodal content into ID construction. Images capture appearance-level cues, whereas texts convey descriptive semantics, making them complementary sources for constructing more expressive semantic IDs.

\begin{figure}[!t]
    \centering
    \includegraphics[width=\columnwidth]{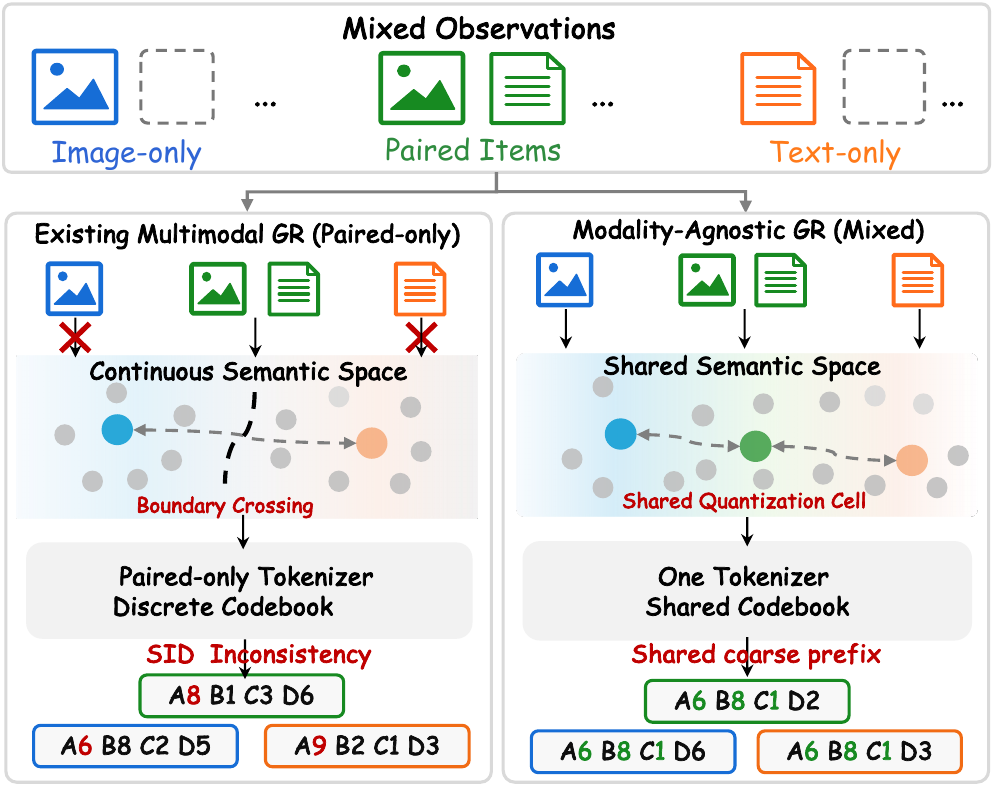}
    
\caption{
Existing multimodal GR methods depend on paired image--text data to learn semantic IDs. However, such paired data is difficult to obtain and maintain in real catalogs, while many useful images and texts are available without clean item-level pairing. UnpairGR learns a shared tokenizer from both paired and unpaired data, producing compatible semantic IDs for a single generative recommender.
}
    \label{fig:intro-motivation}
\end{figure}

Despite these advantages, existing multimodal GR methods usually assume that the image and text used for semantic-ID learning are available as item-level pairs. In real recommendation systems, however, visual and textual information is often generated through different operational pipelines. For example, product images may be uploaded by sellers, whereas titles and descriptions may be maintained through separate catalog-management processes. Since these modalities are collected and updated independently, their coverage and quality can become uneven across items: some items may have reliable images but noisy or missing textual descriptions, while others may contain detailed textual metadata but lack usable images. Even when both modalities are present, catalog evolution still calls for continuous filtering and synchronization to preserve clean item-level image--text pairs~\cite{multimodalrecsurvey}.

Existing methods can only use multimodal information when it appears in paired form. Useful images and useful texts that are not paired with each other cannot directly participate in learning the same semantic-ID tokenizer. This is undesirable because unpaired data still carries valuable item semantics: an image can reveal visual similarity, and a text description can reveal category, function, or usage semantics. The limitation is therefore not the lack of multimodal information, but the requirement that such information must be perfectly paired before it can be used. This motivates our central question:

\noindent\fcolorbox{gray!60}{gray!8}{%
\parbox{\dimexpr\columnwidth-2\fboxsep-2\fboxrule\relax}{%
\vspace{5pt}
\centering
\textbf{Can multimodal GR directly learn expressive semantic IDs from unpaired visual and textual data?}
\vspace{5pt}
}}

However, exploiting unpaired data is particularly challenging in GR, because the final supervision is imposed on discrete semantic IDs rather than continuous embeddings. While standard multimodal representation learning can align visual and textual signals in a shared continuous space, GR must further quantize item representations into discrete IDs before training the autoregressive recommender~\cite{oord2017vqvae,lee2022rqvae,hou2023vqrec}. This quantization makes semantic-ID learning sensitive to representation perturbations: a small shift in the latent space may cross a quantization boundary and produce a different token assignment~\cite{zhu2024cost}.
Such sensitivity is amplified when paired, image-only, and text-only data are encoded through separate transformations or modality-specific codebooks. Semantically related items may remain close in continuous feature space but still be assigned incompatible discrete IDs after quantization. Consequently, unpaired modalities cannot be simply incorporated through independent alignment modules; they must be used to shape the same discrete vocabulary as paired data. The key challenge is therefore to let paired and unpaired visual/textual evidence jointly define a unified semantic-ID space for GR.

A seemingly direct solution is to train a semantic-ID tokenizer jointly on paired, image-only, and text-only observations. As illustrated in Figure~\ref{fig:intro-motivation}, the goal is not to construct separate modality-specific identifier systems or to recover the missing modality, but to map heterogeneous observations into the same semantic space and quantize them with a shared discrete vocabulary. However, mixed supervision alone is insufficient. If different observation conditions are processed by separate semantic transformations, even small representation shifts may cross quantization boundaries and lead to incompatible identifier sequences. Effectively incorporating unpaired data therefore requires paired and unimodal observations to share both the semantic transformation preceding quantization and the residual codebooks used for discrete assignment within a unified tokenizer.

To realize this principle, we propose \textbf{UnpairGR}, short for \textbf{Unpair}ed Modality-Agnostic \textbf{G}enerative \textbf{R}ecommendation. UnpairGR confines modality-specific processing to lightweight input projections, while sharing the subsequent Transformer and residual codebooks across paired, image-only, and text-only observations. Paired image--text observations establish a reliability-guided cross-modal semantic consensus, whereas unpaired images and texts directly update the same representation model and discrete vocabulary. In this way, all available visual and textual evidence contributes to a unified semantic-ID space. After training, the tokenizer is frozen to provide stationary semantic-ID targets for a single autoregressive recommender, without feature imputation, modality-specific codebooks, or fallback mappings.

{\parfillskip=0pt plus 0.5\columnwidth
Extensive experiments on three benchmark datasets demonstrate UnpairGR's effectiveness across diverse settings. Semantic-ID diagnostics show that UnpairGR preserves compatible symbolic neighborhoods across conditions, while cold-start, source-quality, efficiency, and ablation analyses validate the benefits of unpaired supervision, shared representation learning, and residual quantization.\par}

Our main contributions are summarized as follows:

\begin{itemize}
    \item We identify the pairing bottleneck in multimodal GR and formulate learning from paired and unpaired observations as a unified  semantic-ID space learning problem.
  \item We propose Unpaired Modality-Agnostic Generative Recommendation, which enables paired, image-only, and text-only observations to jointly optimize a shared tokenization pathway and residual codebooks, allowing modality evidence to support a single generative recommender.

\item We conduct extensive experiments on three widely used recommendation datasets, demonstrating that UnpairGR consistently outperforms state-of-the-art GR models.
\end{itemize}

\section{Related Work}

In this section, we review representative studies on sequential and generative recommendation, focusing on multimodal item modeling and semantic ID learning.

\subsection{Sequential and Multimodal Recommendation}

Sequential recommendation aims to capture the evolution of user preferences from interaction sequences. Early approaches, such as GRU4Rec and NARM, employ recurrent neural networks to model short-term behavioral transitions and sequential dependencies~\cite{hidasi2016gru4rec,li2017narm}. Attention-based methods improve modeling of long-range user interests. SASRec adopts self-attention to identify informative interactions in user histories, while BERT4Rec introduces bidirectional sequence modeling through a Cloze prediction objective~\cite{kang2018sasrec,sun2019bert4rec}. FDSA and S$^3$-Rec respectively introduce feature-level attention and self-supervised pretraining~\cite{zhang2019fdsa,zhou2020s3rec}. More recently, P5 reformulates recommendation tasks within a unified text-to-text framework, demonstrating the potential of pretrained language models for recommendation~\cite{geng2022p5}.
Multimodal recommendation incorporates visual and textual content to complement interaction supervision~\cite{zhou2023freedom,tao2023slmrec}. MMGCN propagates modality-specific information over user--item graphs to improve preference modeling~\cite{wei2019mmgcn}. MISSRec learns transferable multimodal sequence representations through modality-aware pretraining~\cite{wang2023missrec}, while VIP5 extends prompt-based recommendation to multimodal inputs within a unified generative framework~\cite{geng2023vip5}. These methods demonstrate the value of multimodal content for preference modeling~\cite{zhou2023bm3,zhou2025lvlm,wei2023mmssl,wei2023lightgt}, but primarily operate on continuous representations rather than discrete item vocabularies.

\subsection{Generative Recommendation}

Generative Recommendation (GR) represents items with discrete semantic IDs and predicts the next item autoregressively~\cite{yang2025cobra,zhai2024actions}. TIGER establishes this paradigm through hierarchical residual quantization~\cite{rajput2023tiger}. Subsequent methods improve identifier learning by incorporating collaborative semantics and enhancing codebook utilization, including LC-Rec and LETTER~\cite{zheng2024lcrec,wang2024letter}. Recent methods study end-to-end tokenization~\cite{liu2025etegrec}, multi-identifier pretraining~\cite{zheng2025mtgrec}, and order-agnostic IDs~\cite{lin2025setrec}. Parallel or long IDs~\cite{hou2025rpg,li2026lsig} and differentiable assignments~\cite{fu2026diger} broaden identifier capacity. Tokenizer analyses examine design choices and scaling behavior~\cite{ju2025grid,hou2025tokenization,liu2026scaling}.
Recent studies further extend GR with multimodal information. MMGRec integrates multimodal content and collaborative signals through graph-based residual quantization~\cite{liu2024mmgrec}, while MQL4GRec constructs a unified quantized language for multimodal and cross-domain transfer~\cite{zhai2025mql4grec}. MACRec strengthens cross-modal interaction during quantization and generation, whereas SynGR promotes cross-modal synergy by suppressing dominant-modality shortcuts~\cite{zhang2026macrec,chen2026syngr}. Recent tokenizers also explore mixture quantization~\cite{xu2026mmq}, multimodal initialization~\cite{wang2025mmesid}, non-uniform visual codes~\cite{wei2026card}, and complementary views~\cite{kim2026mviger}.

However, these methods rely on item-level paired observations for cross-modal supervision~\cite{zhang2026macrec,chen2026syngr,xu2026mmq}. Missing-modality approaches instead use feature reconstruction or alignment in continuous spaces~\cite{kim2025dgmrec,wang2023shaspec,lee2023missingprompt,dai2025unbiasedmissing}, without preserving discrete-ID compatibility after quantization. UnpairGR overcomes this limitation by jointly learning from paired, image-only, and text-only observations through a shared Transformer and residual codebooks, yielding a unified semantic-ID space for recommendation.
\begin{figure*}[!t]
\centering
\includegraphics[width=\textwidth]{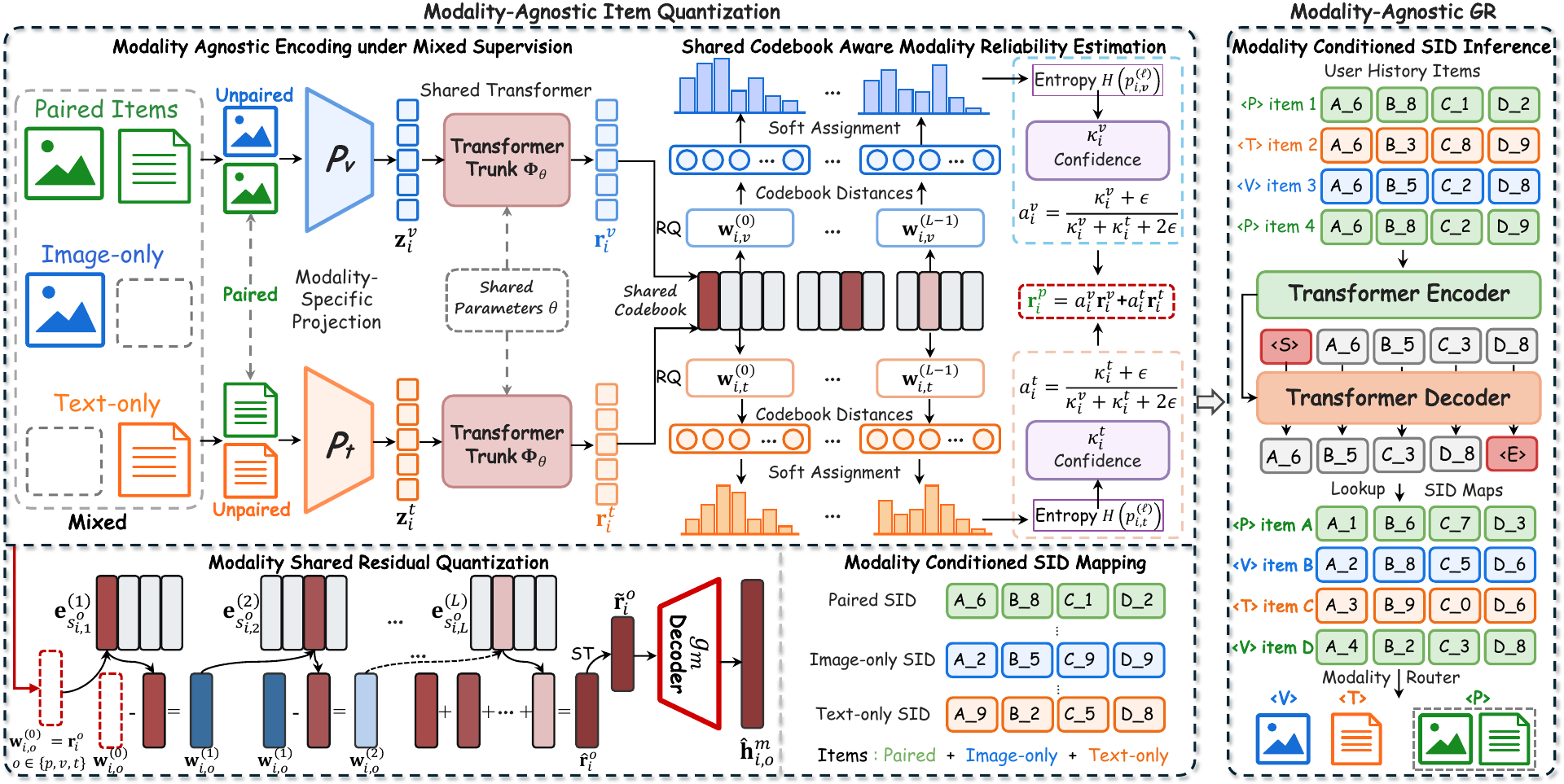}
\caption{Overview of UnpairGR. Paired and unpaired data update the same tokenizer through thin modality-specific projections, a shared Transformer trunk, and shared residual codebooks, yielding semantic IDs in one shared discrete space.}
\label{fig:UnpairGR}
\end{figure*}

\section{Methodology}
\label{sec:method}
Unpaired Modality-Agnostic Generative Recommendation (UnpairGR) learns semantic IDs from paired, image-only, and text-only item collections, denoted by $\mathcal{D}_{p}$, $\mathcal{D}_{v}$, and $\mathcal{D}_{t}$. Rather than maintaining separate tokenizers, it maps all observations into a shared representation space and quantizes them with common residual codebooks, yielding IDs for a single autoregressive recommender. As shown in Figure~\ref{fig:UnpairGR}, the tokenizer learns shared representations, forms a cross-modal consensus for paired items, and applies the same residual quantizer to all observation types. After training, it is frozen to provide stationary targets for generative recommendation.

\subsection{Modality-Agnostic Item Quantization}

The objective of modality-agnostic item quantization is to construct a common semantic-ID space that can be independently accessed through visual, textual, or paired observations. This requires sharing not only the discrete codes but also the semantic transformation preceding quantization. Otherwise, unimodal observations may update isolated modality-specific representations and produce identifiers that are incompatible with those derived from paired inputs. UnpairGR therefore confines modality-specific processing to lightweight input projections, while sharing the subsequent representation model and residual codebooks across all observation conditions.
\subsubsection{Shared Representation across Modalities}

For an item $i$, frozen modality encoders extract visual and textual features
$\mathbf{h}_{i}^{v}\in\mathbb{R}^{d_v}$ and
$\mathbf{h}_{i}^{t}\in\mathbb{R}^{d_t}$.
Freezing the encoders stabilizes modality distributions during semantic-ID learning and prevents the shared quantizer from tracking independently evolving representation spaces~\cite{radford2021clip}.
Because modalities differ in dimensionality and statistics, each feature is mapped by a lightweight projection $P_m$ and then processed by a shared Transformer $\Phi_{\theta}$:
\begin{equation}
    \mathbf{z}_{i}^{m}
    =
    P_m\!\left(\mathbf{h}_{i}^{m}\right),
    \qquad
    \mathbf{r}_{i}^{m}
    =
    \Phi_{\theta}\!\left(\mathbf{z}_{i}^{m}\right),
    \quad m\in\{v,t\}.
    \label{eq:shared_representation}
\end{equation}
The projections $P_v$ and $P_t$ absorb modality-specific discrepancies, while $\Phi_{\theta}$ provides a shared semantic transformation before quantization. Consequently, image-only and text-only observations can update the same parameters without item-level counterparts, allowing unimodal evidence to more effectively refine a coherent semantic space shared across heterogeneous observation conditions, rather than forming separate modality-specific representations.
For paired items, $\mathbf{r}_{i}^{v}$ and $\mathbf{r}_{i}^{t}$ provide complementary views. Since direct averaging assumes equal reliability, UnpairGR instead forms a cross-modal consensus by estimating each representation's confidence in the shared quantization space.

\subsubsection{Cross-Modal Semantic Consensus}
For a paired item, visual and textual representations may provide unequal semantic evidence, so direct averaging can let an ambiguous modality distort semantic ID construction. UnpairGR instead estimates modality reliability in a principled manner from assignment uncertainty in the shared quantization space.

Let $\mathcal{C}^{(\ell)}=\{\mathbf{e}_{k}^{(\ell)}\}_{k=1}^{K}$ be the shared codebook at residual level $\ell$, and let $\mathbf{w}_{i,m}^{(\ell-1)}$ be the residual for modality $m\in\{v,t\}$. Its soft assignment over the shared codebook is
\begin{equation}
p_{i,m}^{(\ell)}(k)
=
\frac{
\exp\!\left(
-\left\|
\mathbf{w}_{i,m}^{(\ell-1)}
-
\mathbf{e}_{k}^{(\ell)}
\right\|_{2}^{2}/\tau
\right)
}{
\sum_{j=1}^{K}
\exp\!\left(
-\left\|
\mathbf{w}_{i,m}^{(\ell-1)}
-
\mathbf{e}_{j}^{(\ell)}
\right\|_{2}^{2}/\tau
\right)
},
\label{eq:soft_assignment}
\end{equation}
where $\tau$ controls assignment sharpness. We define modality reliability using the average normalized entropy:
\begin{equation}
\kappa_i^m
=
1-
\frac{1}{L}
\sum_{\ell=1}^{L}
\frac{
H\!\left(p_{i,m}^{(\ell)}\right)
}{
\log K
},
\label{eq:modality_reliability}
\end{equation}
where $H(p)=-\sum_k p(k)\log p(k)$. Lower entropy indicates more confident localization in the shared code space and thus higher reliability.
We normalize the scores as
$a_i^m=(\kappa_i^m+\epsilon)/(\kappa_i^v+\kappa_i^t+2\epsilon)$,
where $\epsilon>0$, and construct
$\mathbf{r}_i^p=a_i^v\mathbf{r}_i^v+a_i^t\mathbf{r}_i^t$.
Its assignment $p_{i,p}^{(\ell)}$ follows Eq.~\eqref{eq:soft_assignment} using the paired residual $\mathbf{w}_{i,p}^{(\ell-1)}$.
For unimodal items, the available representation is used directly without fusion.

\subsubsection{Modality-Shared Residual Quantization}

Given an item representation $\mathbf{r}_i^o$ under observation condition $o\in\{p,v,t\}$, UnpairGR discretizes all observation types using the same $L$-level residual codebook
$\{\mathcal{C}^{(\ell)}\}_{\ell=1}^{L}$, where
$\mathcal{C}^{(\ell)}=\{\mathbf{e}_k^{(\ell)}\}_{k=1}^{K}$.
Starting from $\mathbf{w}_{i,o}^{(0)}=\mathbf{r}_i^o$, residual quantization proceeds as
\begin{equation}
s_{i,\ell}^{o}
=
\underset{k\in[K]}{\arg\min}\,
\left\|
\mathbf{w}_{i,o}^{(\ell-1)}
-
\mathbf{e}_{k}^{(\ell)}
\right\|_{2}^{2}.
\label{eq:code_selection}
\end{equation}

\begin{equation}
\mathbf{w}_{i,o}^{(\ell)}
=
\mathbf{w}_{i,o}^{(\ell-1)}
-
\mathbf{e}_{s_{i,\ell}^{o}}^{(\ell)}.
\label{eq:residual_update}
\end{equation}
The selected indices form the semantic ID
$\mathbf{s}_i^o=(s_{i,1}^o,\ldots,s_{i,L}^o)$, with quantized representation
$\widehat{\mathbf{r}}_i^o=\sum_{\ell=1}^{L}\mathbf{e}_{s_{i,\ell}^{o}}^{(\ell)}$.
Sharing residual codebooks across paired, image-only, and text-only observations yields coarse-to-fine IDs in a common discrete space, allowing unimodal data to refine the same codes used by paired items. We apply nearest-code assignment uniformly and resolve collisions by deterministic suffix reassignment at the final levels.

\subsection{Unpaired Semantic ID Learning}

A shared representation model and residual codebooks establish a common quantization space, but do not guarantee effective use of unimodal observations. UnpairGR therefore jointly learns from paired, image-only, and text-only data. Paired items provide cross-modal correspondence, while unimodal items refine shared representations and codes directly from available content, without modality completion.

\subsubsection{Mixed-Supervision Quantization}

UnpairGR jointly learns from paired, image-only, and text-only observations through a unified quantization objective. Let $o\in\{p,v,t\}$ denote the observation condition, and let $\mathcal{M}_{p}=\{v,t\}$, $\mathcal{M}_{v}=\{v\}$, and $\mathcal{M}_{t}=\{t\}$ specify the available modalities. Accordingly, paired observations are supervised to reconstruct both modalities, whereas unimodal observations reconstruct only their observed modality.
Let $\ell_{\mathrm{RQ}}(\mathbf{r})$ denote the residual quantization loss defined in the Supplementary Appendix, $g_m$ a training-only reconstruction head for modality $m$, and $\operatorname{sg}[\cdot]$ the stop-gradient operator. For observation condition $o$, we employ the straight-through representation $\widetilde{\mathbf{r}}_{i}^{o}=\mathbf{r}_{i}^{o}+\operatorname{sg}[\widehat{\mathbf{r}}_{i}^{o}-\mathbf{r}_{i}^{o}]$. For each available modality $m\in\mathcal{M}_{o}$, we denote the reconstructed modality feature by $\widehat{\mathbf{h}}_{i,o}^{m}:=g_m(\widetilde{\mathbf{r}}_{i}^{o})$ and define the observation-specific reconstruction loss as $\mathcal{R}_{i}^{o}=\sum_{m\in\mathcal{M}_{o}}\|\widehat{\mathbf{h}}_{i,o}^{m}-\mathbf{h}_{i}^{m}\|_{2}^{2}$. The mixed-supervision objective is
\begin{equation}
\mathcal{L}_{\mathrm{mix}}
=
\sum_{o\in\{p,v,t\}}
\mathbb{E}_{i\sim\mathcal{D}_{o}}
\left[
\ell_{\mathrm{RQ}}(\mathbf{r}_{i}^{o})
+
\mathcal{R}_{i}^{o}
\right].
\label{eq:mixed_supervision}
\end{equation}
Although image-only and text-only observations lack item-level cross-modal correspondence, they directly optimize the same Transformer and residual codebooks as paired observations. Unimodal evidence therefore contributes directly to semantic ID learning rather than serving as auxiliary feature completion. During training, we draw balanced samples from $\mathcal{D}_{p}$, $\mathcal{D}_{v}$, and $\mathcal{D}_{t}$ to prevent any observation type from dominating code assignment.

\subsubsection{Cross-Observation Semantic Regularization}

Shared parameters encourage different observations to inhabit a common space, but do not guarantee that they produce compatible semantic IDs. UnpairGR therefore enforces consistency both before and after quantization.
For each paired item, we align its visual and textual representations through
$\mathcal{L}_{\mathrm{align}}
=
\mathbb{E}_{i\sim\mathcal{D}_{p}}
\|\mathbf{r}_{i}^{v}-\mathbf{r}_{i}^{t}\|_{2}^{2}$.
We further define the assignment discrepancy at residual level $\ell$ as
$\Delta_i^{(\ell)}
=
\sum_{m\in\{v,t\}}
D_{\mathrm{JS}}
(p_{i,p}^{(\ell)}\|p_{i,m}^{(\ell)})$,
where $D_{\mathrm{JS}}$ denotes the Jensen--Shannon divergence, a symmetric measure of discrepancy between two probability distributions. The cross-observation consistency objective is
\begin{equation}
\mathcal{L}_{\mathrm{cons}}
=
\mathbb{E}_{i\sim\mathcal{D}_{p}}
\left[
\frac{1}{L}
\sum_{\ell=1}^{L}
\omega_{\ell}\Delta_i^{(\ell)}
\right].
\label{eq:cross_observation_consistency}
\end{equation}
Here, $p_{i,p}^{(\ell)}$ denotes the soft assignment of the paired consensus, while $p_{i,m}^{(\ell)}$ denotes its visual or textual counterpart. The coefficient $\omega_{\ell}$ emphasizes earlier residual levels encoding coarser semantics. While $\mathcal{L}_{\mathrm{align}}$ reduces cross-modal discrepancy before quantization, $\mathcal{L}_{\mathrm{cons}}$ preserves assignment compatibility after quantization. Together, they prevent observations of the same item from producing inconsistent semantic IDs.

\subsubsection{Balanced Code Allocation}

Mixed supervision may concentrate assignments on a small subset of codes. To prevent code collapse, we regularize mini-batch code usage. Let $\bar{p}^{(\ell)}$ be the mean soft assignment at residual level $\ell$, and let $u_K$ denote the uniform distribution over $K$ codes. We define
\begin{equation}
\mathcal{L}_{\mathrm{use}}
=
\frac{1}{L}
\sum_{\ell=1}^{L}
\operatorname{KL}
\left(
\bar{p}^{(\ell)}
\,\|\, 
u_K
\right).
\label{eq:code_utilization}
\end{equation}
This objective encourages broad coverage of the shared codebooks without imposing uniform assignments on individual items.  The complete tokenization objective is
\begin{equation}
\mathcal{L}_{\mathrm{tok}}
=
\mathcal{L}_{\mathrm{mix}}
+
\lambda\mathcal{L}_{\mathrm{align}}
+
\eta\mathcal{L}_{\mathrm{cons}}
+
\xi\mathcal{L}_{\mathrm{use}}.
\label{eq:tokenization_objective}
\end{equation}
We sample balanced mini-batches from $\mathcal{D}_{p}$, $\mathcal{D}_{v}$, and $\mathcal{D}_{t}$ to prevent any observation type from dominating optimization.

Training proceeds in two stages. First, UnpairGR minimizes $\mathcal{L}_{\mathrm{tok}}$ and freezes the modality projections, shared Transformer, and residual codebooks to construct semantic IDs. Second, the autoregressive recommender is trained on these fixed IDs. Paired IDs are used when both modalities are available; otherwise, the corresponding unimodal IDs are used.
\begin{table*}[!t]
  \centering
  \scriptsize
  \setlength{\tabcolsep}{2.15pt}
  \renewcommand{\arraystretch}{1.06}
  \begin{tabular}{@{}l*{15}{c}@{}}
    \toprule
    \multirow{2}{*}{Model / Datasets} & \multicolumn{5}{c}{Arts} & \multicolumn{5}{c}{Games} & \multicolumn{5}{c}{Instruments} \\
    \cmidrule(lr){2-6} \cmidrule(lr){7-11} \cmidrule(lr){12-16}
    & HR@1 & HR@5 & HR@10 & N@5 & N@10 & HR@1 & HR@5 & HR@10 & N@5 & N@10 & HR@1 & HR@5 & HR@10 & N@5 & N@10 \\
    \midrule
    GRU4Rec \UnpairGRVenue{(ICLR'16)} & 0.0365 & 0.0817 & 0.1088 & 0.0602 & 0.0690 & 0.0140 & 0.0544 & 0.0895 & 0.0341 & 0.0453 & 0.0566 & 0.0975 & 0.1207 & 0.0783 & 0.0857 \\
    SASRec \UnpairGRVenue{(ICDM'18)} & 0.0212 & 0.0951 & 0.1250 & 0.0610 & 0.0706 & 0.0069 & 0.0587 & 0.0985 & 0.0333 & 0.0461 & 0.0318 & 0.0946 & 0.1233 & 0.0654 & 0.0746 \\
    BERT4Rec \UnpairGRVenue{(CIKM'19)} & 0.0289 & 0.0697 & 0.0922 & 0.0502 & 0.0575 & 0.0115 & 0.0426 & 0.0725 & 0.0270 & 0.0366 & 0.0450 & 0.0856 & 0.1081 & 0.0667 & 0.0739 \\
    FDSA \UnpairGRVenue{(IJCAI'19)} & 0.0380 & 0.0832 & 0.1190 & 0.0583 & 0.0695 & 0.0163 & 0.0614 & 0.0988 & 0.0389 & 0.0509 & 0.0530 & 0.0987 & 0.1249 & 0.0775 & 0.0859 \\
    S\textsuperscript{3}-Rec \UnpairGRVenue{(CIKM'20)} & 0.0172 & 0.0739 & 0.1030 & 0.0511 & 0.0630 & 0.0136 & 0.0527 & 0.0903 & 0.0351 & 0.0468 & 0.0339 & 0.0937 & 0.1123 & 0.0693 & 0.0743 \\
    P5-CID \UnpairGRVenue{(RecSys'22)} & 0.0421 & 0.0713 & 0.0994 & 0.0607 & 0.0662 & 0.0169 & 0.0532 & 0.0824 & 0.0331 & 0.0454 & 0.0512 & 0.0839 & 0.1119 & 0.0678 & 0.0704 \\
    VQ-Rec \UnpairGRVenue{(WWW'23)} & 0.0408 & 0.1038 & 0.1386 & 0.0732 & 0.0844 & 0.0075 & 0.0408 & 0.0679 & 0.0242 & 0.0329 & 0.0502 & 0.1062 & 0.1357 & 0.0796 & 0.0891 \\
    MISSRec \UnpairGRVenue{(MM'23)} & 0.0479 & 0.1021 & 0.1321 & 0.0699 & 0.0815 & 0.0201 & 0.0674 & 0.1048 & 0.0385 & 0.0499 & 0.0723 & 0.1089 & 0.1361 & 0.0797 & 0.0880 \\
    VIP5 \UnpairGRVenue{(EMNLP'23)} & 0.0474 & 0.0704 & 0.0859 & 0.0586 & 0.0635 & 0.0173 & 0.0480 & 0.0758 & 0.0328 & 0.0418 & 0.0737 & 0.0892 & 0.1071 & 0.0815 & 0.0872 \\
    TIGER \UnpairGRVenue{(NeurIPS'23)} & 0.0532 & 0.0894 & 0.1167 & 0.0718 & 0.0806 & 0.0166 & 0.0523 & 0.0857 & 0.0345 & 0.0453 & 0.0754 & 0.1007 & 0.1221 & 0.0882 & 0.0950 \\
    MQL4GRec \UnpairGRVenue{(ICLR'25)} & 0.0672 & 0.1037 & 0.1327 & 0.0857 & 0.0950 & 0.0203 & 0.0637 & 0.1033 & 0.0421 & 0.0548 & 0.0833 & 0.1115 & 0.1375 & 0.0977 & 0.1060 \\
    MACRec \UnpairGRVenue{(AAAI'26)} & 0.0685 & 0.1046 & 0.1329 & 0.0868 & 0.0953 & 0.0208 & 0.0671 & 0.1078 & 0.0435 & 0.0565 & 0.0819 & 0.1110 & 0.1363 & 0.0965 & 0.1046 \\
    SynGR \UnpairGRVenue{(ICML'26)} & \UnpairGRSecond{0.0713} & \UnpairGRSecond{0.1145} & \UnpairGRSecond{0.1449} & \UnpairGRSecond{0.0919} & \UnpairGRSecond{0.1010} & \UnpairGRSecond{0.0245} & \UnpairGRSecond{0.0702} & \UnpairGRSecond{0.1092} & \UnpairGRSecond{0.0471} & \UnpairGRSecond{0.0596} & \UnpairGRSecond{0.0850} & \UnpairGRSecond{0.1321} & \UnpairGRSecond{0.1768} & \UnpairGRSecond{0.1045} & \UnpairGRSecond{0.1189} \\
    \midrule
    UnpairGR (Ours) & \UnpairGRBest{0.0876} & \UnpairGRBest{0.1262} & \UnpairGRBest{0.1550} & \UnpairGRBest{0.1075} & \UnpairGRBest{0.1168} & \UnpairGRBest{0.0287} & \UnpairGRBest{0.0753} & \UnpairGRBest{0.1113} & \UnpairGRBest{0.0523} & \UnpairGRBest{0.0638} & \UnpairGRBest{0.0954} & \UnpairGRBest{0.1452} & \UnpairGRBest{0.1769} & \UnpairGRBest{0.1214} & \UnpairGRBest{0.1315} \\
    Improv. (\%) & \UnpairGRGain{+22.86\%} & \UnpairGRGain{+10.22\%} & \UnpairGRGain{+6.97\%} & \UnpairGRGain{+16.97\%} & \UnpairGRGain{+15.64\%} & \UnpairGRGain{+17.14\%} & \UnpairGRGain{+7.26\%} & \UnpairGRGain{+1.92\%} & \UnpairGRGain{+11.04\%} & \UnpairGRGain{+7.05\%} & \UnpairGRGain{+12.24\%} & \UnpairGRGain{+9.92\%} & \UnpairGRGain{+0.06\%} & \UnpairGRGain{+16.17\%} & \UnpairGRGain{+10.60\%} \\
    \bottomrule
  \end{tabular}
\caption{
Overall recommendation performance under fully observed inputs.
\textit{N} denotes NDCG.
The best and second-best are highlighted in bold and underlined, respectively.
\textit{Improv.} reports the relative gain of UnpairGR over the strongest baseline, SynGR.
}
\label{tab:main-results}
\end{table*}
\begin{table*}[!t]
  \centering
    \scriptsize
  \setlength{\tabcolsep}{2.15pt}
  \renewcommand{\arraystretch}{1.06}
  \begin{tabular*}{\textwidth}{@{\extracolsep{\fill}}ll*{13}{c}@{}}
    \toprule
    \multirow{2}{*}{Dataset} & \multirow{2}{*}{Method}
      & \multicolumn{1}{c}{Full}
      & \multicolumn{3}{c}{Missing 25\%}
      & \multicolumn{3}{c}{Missing 50\%}
      & \multicolumn{3}{c}{Missing 75\%}
      & \multicolumn{3}{c}{Missing 100\%} \\
    \cmidrule(lr){3-3}\cmidrule(lr){4-6}\cmidrule(lr){7-9}\cmidrule(lr){10-12}\cmidrule(l){13-15}
    & & All
      & TO
      & IO
      & Random
      & TO
      & IO
      & Random
      & TO
      & IO
      & Random
      & TO
      & IO
      & Random \\
    \midrule
    \multirow{5}{*}{Arts}
        & MQL4GRec     & 0.1327 & \UnpairGRSecond{0.1251} & 0.1086 & \UnpairGRSecond{0.1133} & \UnpairGRSecond{0.1256} & \UnpairGRSecond{0.1027} & \UnpairGRSecond{0.1095} & \UnpairGRSecond{0.1257} & \UnpairGRSecond{0.1083} & \UnpairGRSecond{0.1045} & \UnpairGRSecond{0.1253} & \UnpairGRSecond{0.1234} & 0.1050 \\
      & MACRec      & 0.1329 & 0.1072 & 0.0979 & 0.0958 & 0.0977 & 0.0931 & 0.0969 & 0.0998 & 0.0914 & 0.0879 & 0.0911 & 0.0866 & 0.0888 \\

      & SynGR     & \UnpairGRSecond{0.1449} & 0.1193 & \UnpairGRSecond{0.1194} & 0.1015 & 0.1192 & 0.0511 & 0.0856 & 0.1178 & 0.0515 & 0.0886 & 0.0675 & 0.0755 & \UnpairGRSecond{0.1171} \\
      & UnpairGR    & \UnpairGRBest{0.1550} & \UnpairGRBest{0.1281} & \UnpairGRBest{0.1239} & \UnpairGRBest{0.1202} & \UnpairGRBest{0.1277} & \UnpairGRBest{0.1169} & \UnpairGRBest{0.1180} & \UnpairGRBest{0.1273} & \UnpairGRBest{0.1170} & \UnpairGRBest{0.1109} & \UnpairGRBest{0.1321} & \UnpairGRBest{0.1281} & \UnpairGRBest{0.1202} \\
      & Improv. (\%) & \UnpairGRGain{+6.97\%} & \UnpairGRGain{+2.40\%} & \UnpairGRGain{+3.77\%} & \UnpairGRGain{+6.09\%} & \UnpairGRGain{+1.67\%} & \UnpairGRGain{+13.83\%} & \UnpairGRGain{+7.76\%} & \UnpairGRGain{+1.27\%} & \UnpairGRGain{+8.03\%} & \UnpairGRGain{+6.12\%} & \UnpairGRGain{+5.43\%} & \UnpairGRGain{+3.81\%} & \UnpairGRGain{+2.65\%} \\
    \midrule
    \multirow{5}{*}{Games}
        & MQL4GRec     & 0.1033 & \UnpairGRSecond{0.0935} & \UnpairGRSecond{0.0775} & \UnpairGRSecond{0.0854} & \UnpairGRSecond{0.0930} & 0.0739 & \UnpairGRSecond{0.0746} & \UnpairGRSecond{0.0931} & \UnpairGRSecond{0.0839} & 0.0717 & \UnpairGRSecond{0.0932} & \UnpairGRSecond{0.0942} & 0.0699 \\
      & MACRec      & 0.1078 & 0.0689 & 0.0648 & 0.0568 & 0.0727 & 0.0643 & 0.0670 & 0.0659 & 0.0665 & 0.0689 & 0.0659 & 0.0623 & 0.0674 \\

      & SynGR     & \UnpairGRSecond{0.1092} & 0.0907 & 0.0364 & 0.0704 & 0.0895 & \UnpairGRSecond{0.0906} & 0.0635 & 0.0443 & 0.0426 & \UnpairGRSecond{0.0737} & 0.0436 & 0.0553 & \UnpairGRSecond{0.0894} \\

      & UnpairGR    & \UnpairGRBest{0.1113} & \UnpairGRBest{0.0948} & \UnpairGRBest{0.0904} & \UnpairGRBest{0.0925} & \UnpairGRBest{0.0931} & \UnpairGRBest{0.0937} & \UnpairGRBest{0.0969} & \UnpairGRBest{0.0947} & \UnpairGRBest{0.0865} & \UnpairGRBest{0.0969} & \UnpairGRBest{0.0967} & \UnpairGRBest{0.0952} & \UnpairGRBest{0.0969} \\
      & Improv. (\%) & \UnpairGRGain{+1.92\%} & \UnpairGRGain{+1.39\%} & \UnpairGRGain{+16.65\%} & \UnpairGRGain{+8.31\%} & \UnpairGRGain{+0.11\%} & \UnpairGRGain{+3.42\%} & \UnpairGRGain{+29.89\%} & \UnpairGRGain{+1.72\%} & \UnpairGRGain{+3.10\%} & \UnpairGRGain{+31.48\%} & \UnpairGRGain{+3.76\%} & \UnpairGRGain{+1.06\%} & \UnpairGRGain{+8.39\%} \\
    \midrule
    \multirow{5}{*}{Instruments}
        & MQL4GRec     & 0.1375 & \UnpairGRSecond{0.1342} & \UnpairGRSecond{0.1211} & \UnpairGRSecond{0.1227} & \UnpairGRSecond{0.1328} & \UnpairGRSecond{0.1238} & \UnpairGRSecond{0.1197} & \UnpairGRSecond{0.1346} & \UnpairGRSecond{0.1235} & \UnpairGRSecond{0.1219} & \UnpairGRSecond{0.1353} & \UnpairGRSecond{0.1344} & 0.1127 \\
      & MACRec      & 0.1363 & 0.1126 & 0.1184 & 0.1106 & 0.1156 & 0.1149 & 0.1149 & 0.1148 & 0.1137 & 0.1124 & 0.1082 & 0.1110 & \UnpairGRSecond{0.1135} \\
      & SynGR     & \UnpairGRSecond{0.1768} & 0.1311 & 0.0640 & 0.0953 & 0.1293 & 0.0828 & 0.0821 & 0.1308 & 0.1022 & 0.1040 & 0.0913 & 0.1088 & 0.0676 \\
      & UnpairGR    & \UnpairGRBest{0.1769} & \UnpairGRBest{0.1378} & \UnpairGRBest{0.1359} & \UnpairGRBest{0.1236} & \UnpairGRBest{0.1352} & \UnpairGRBest{0.1244} & \UnpairGRBest{0.1280} & \UnpairGRBest{0.1354} & \UnpairGRBest{0.1268} & \UnpairGRBest{0.1229} & \UnpairGRBest{0.1358} & \UnpairGRBest{0.1398} & \UnpairGRBest{0.1209} \\
      & Improv. (\%) & \UnpairGRGain{+0.06\%} & \UnpairGRGain{+2.68\%} & \UnpairGRGain{+12.22\%} & \UnpairGRGain{+0.73\%} & \UnpairGRGain{+1.81\%} & \UnpairGRGain{+0.48\%} & \UnpairGRGain{+6.93\%} & \UnpairGRGain{+0.59\%} & \UnpairGRGain{+2.67\%} & \UnpairGRGain{+0.82\%} & \UnpairGRGain{+0.37\%} & \UnpairGRGain{+4.02\%} & \UnpairGRGain{+6.52\%} \\
    \bottomrule
  \end{tabular*}

\caption{
Recommendation robustness under matched train–test modality incompleteness, with three multimodal GR baselines adapted for mixed observations. HR@10 is reported for text-only (TO), image-only (IO), and random single-modality (Random) conditions. \textit{Improv.} is the gain over the strongest baseline; NDCG@10 is in Supplementary Table~A1.
}
\label{tab:train-time-missing-hr}
\end{table*}

\subsubsection{Semantic ID Generation under Heterogeneous Observations}

After tokenization, all items are represented by semantic IDs within one semantic-ID space, regardless of their observation conditions. For each interacted item $i$, paired observations use the cross-modal consensus, whereas image-only and text-only observations use their corresponding unimodal representations. The resulting semantic ID is denoted by
$\mathbf{s}_i=(s_{i,1},\ldots,s_{i,L})$.
In Figure~\ref{fig:UnpairGR}, $\langle P\rangle$, $\langle V\rangle$, and $\langle T\rangle$ denote paired, image-only, and text-only tokenization routes, respectively, and are not included in the input sequence.
For a user $u$, the interaction history is formed by concatenating the semantic IDs of previously interacted items. Given the encoded history $\mathbf{x}_u$ and the target semantic ID
$\mathbf{y}_u=(y_1,\ldots,y_L)$, the generative recommender is optimized with
\begin{equation}
\mathcal{L}_{\mathrm{rec}}
=
-\sum_{\ell=1}^{L}
\log P_{\psi}
\left(
y_{\ell}
\mid
y_{<\ell},
\mathbf{x}_u
\right).
\label{eq:recommendation_loss}
\end{equation}

At inference, available content determines the tokenization route. Because all routes share the Transformer and residual codebooks, paired and unimodal observations are mapped into a common discrete ID space and handled by a single autoregressive recommender. Constrained beam search then generates valid semantic IDs and retrieves the items, without feature imputation or modality-specific fallback models.

\section{Experiments}
We address six research questions: \textbf{RQ1:} How effective is UnpairGR with complete inputs? \textbf{RQ2:} How robust is UnpairGR when modality incompleteness affects both training and evaluation? \textbf{RQ3:} Do different observations of the same item yield compatible semantic IDs? \textbf{RQ4:} Can it generalize to single-modality cold-start items? \textbf{RQ5:} Which components drive its effectiveness? \textbf{RQ6:} How efficient is it in training and inference? Datasets, baselines, protocols, implementation details, partial-pairing setup, hyperparameter analyses, and case studies are provided in the Supplementary Appendix.

\subsection{Overall Recommendation Performance (RQ1)}
\label{sec:overall_performance}

Table~\ref{tab:main-results} compares UnpairGR with representative sequential, multimodal, and generative recommendation methods under fully observed inputs. UnpairGR achieves the best results across all datasets and metrics, showing that paired and unimodal observations remain beneficial when complete multimodal information is available at inference.
The gains are most pronounced at the highest-ranked positions. Compared with SynGR, UnpairGR improves HR@1 by $22.86\%$ on Arts and $17.14\%$ on Games, and NDCG@5 by $16.17\%$ on Instruments. The gains persist at larger cutoffs, although the margin on Instruments HR@10 is modest. This pattern indicates that UnpairGR more accurately identifies relevant items at the top of the ranking.
UnpairGR also consistently outperforms conventional sequential recommenders and earlier semantic-ID-based generative methods. These results show that unpaired supervision improves both robustness to incomplete observations and semantic ID construction under fully observed inputs. RQ5 examines component contributions.

\subsection{Robustness to Incomplete Observations (RQ2)}
\label{sec:rq2-missing-robustness}
We evaluate UnpairGR under matched modality incompleteness during training and evaluation, reporting HR@10 in Table~\ref{tab:train-time-missing-hr} across text-only (TO), image-only (IO), and random single-modality (Random) conditions at varying missingness ratios. UnpairGR consistently performs best across datasets and conditions, with its advantage persisting as missingness increases. Strong performance under fully unimodal inputs further shows that image-only and text-only observations remain compatible with the shared tokenizer and recommender. Overall, mixed supervision yields a coherent discrete space robust to heterogeneous modality availability.

\begin{table}[!t]
  \centering
  \scriptsize
  \setlength{\tabcolsep}{0.5pt}
  \renewcommand{\arraystretch}{0.88}
  \begin{tabular*}{\columnwidth}{@{\extracolsep{\fill}}l@{\hspace{4pt}}lcccc@{\hspace{5pt}}cccc@{}}
    \toprule
    Dataset & Setting & \multicolumn{4}{c}{Text-only ID} & \multicolumn{4}{c}{Image-only ID} \\
    \cmidrule(lr){3-6} \cmidrule(lr){7-10}
    & & Exact & P1$\uparrow$ & P2$\uparrow$ & Dist.$\downarrow$ & Exact & P1$\uparrow$ & P2$\uparrow$ & Dist.$\downarrow$ \\
    \midrule
    Arts & \textbf{Observed} & \textbf{0.124} & \textbf{0.863} & \textbf{0.549} & \textbf{0.461} & \textbf{0.124} & \textbf{0.860} & \textbf{0.528} & \textbf{0.467} \\
    & Random & 0.000 & 0.070 & 0.002 & 0.976 & 0.000 & 0.064 & 0.002 & 0.978 \\
    Games & \textbf{Observed} & \textbf{0.136} & \textbf{0.877} & \textbf{0.579} & \textbf{0.444} & \textbf{0.115} & \textbf{0.866} & \textbf{0.508} & \textbf{0.488} \\
    & Random & 0.000 & 0.103 & 0.002 & 0.969 & 0.000 & 0.103 & 0.004 & 0.969 \\
    Instruments & \textbf{Observed} & \textbf{0.146} & \textbf{0.889} & \textbf{0.608} & \textbf{0.428} & \textbf{0.126} & \textbf{0.880} & \textbf{0.570} & \textbf{0.452} \\
    & Random & 0.000 & 0.143 & 0.005 & 0.957 & 0.000 & 0.140 & 0.005 & 0.958 \\
    \midrule
    All & \textbf{Observed} & \textbf{0.134} & \textbf{0.875} & \textbf{0.576} & \textbf{0.446} & \textbf{0.120} & \textbf{0.867} & \textbf{0.528} & \textbf{0.474} \\
    & Random & 0.000 & 0.101 & 0.003 & 0.969 & 0.000 & 0.098 & 0.003 & 0.969 \\
    \bottomrule
  \end{tabular*}
  \caption{Cross-observation semantic-ID consistency. Random denotes within-dataset permutations; Dist. is normalized Hamming distance; All is item-weighted.}
  \label{tab:rq3-id-stability}
\end{table}

\subsection{Cross-Observation Semantic-ID Consistency (RQ3)}
To verify that UnpairGR learns a shared discrete vocabulary rather than merely improving accuracy, we test whether observations of the same item yield compatible semantic IDs. Table~\ref{tab:rq3-id-stability} compares text-only and image-only IDs with paired references using exact agreement, prefix agreement, and normalized Hamming distance against random permutations. UnpairGR achieves much higher prefix agreement and lower Hamming distance across all datasets. Although exact matches are less frequent, early codes remain highly consistent, indicating shared coarse-grained semantics with variation mainly in later codes. This pattern is consistent with the coarse-to-fine structure of residual quantization, where earlier codes capture broad semantics and later codes encode finer item-specific details. This directly supports the use of one semantic-ID space for paired and unimodal observations.

\subsection{Single-Modality Cold-Start Generalization (RQ4)}
\label{sec:cold-start-generalization}

\begin{figure}[!tbp]
  \centering
  \includegraphics[width=\columnwidth]{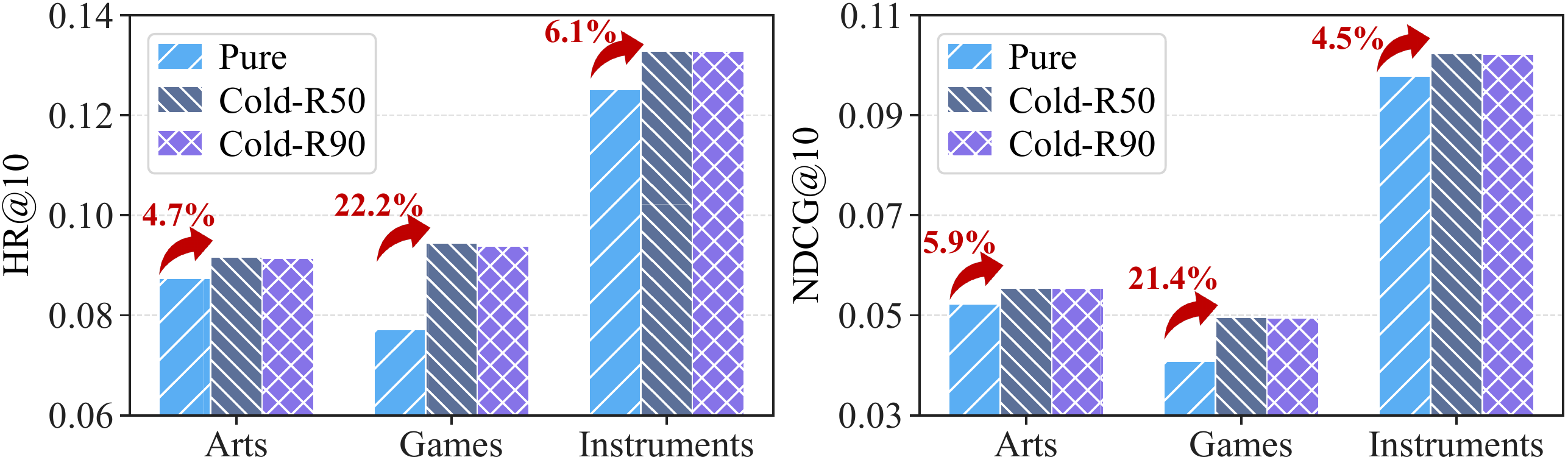}
  \caption{Cold-start mechanism analysis for UnpairGR. HR@10 and NDCG@10 compare the Pure route with validation-selected Cold-R50 and Cold-R90 routes under deterministic target-side single-modality masks.}
  
  \label{fig:rq4-coldstart-mechanism}
\end{figure}

We evaluate whether semantic IDs derived from a single modality can support cold-start recommendation~\cite{zhao2026coins,zheng2026utgrec}. Here, Pure denotes the standard route, while Cold-R50 and Cold-R90 apply deterministic single-modality masking to 50\% and 90\% of target items, respectively. Figure~\ref{fig:rq4-coldstart-mechanism} compares these three routes.
Both cold-start configurations consistently outperform the Pure route across the three datasets, with the largest gains observed on Games. These results show that single-modality semantic IDs remain compatible with the same autoregressive recommender, enabling cold-start recommendation without feature imputation or a separate fallback model.

\subsection{Component Analysis (RQ5)}

\begin{figure}[!tbp]
  \centering
  \includegraphics[width=\columnwidth]{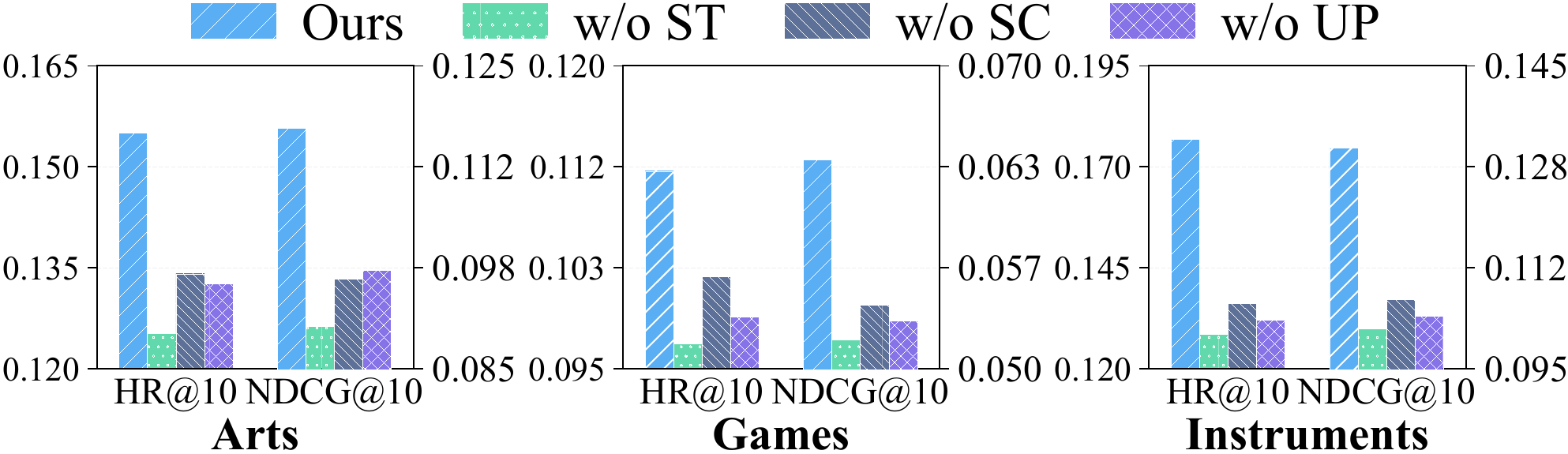}
  \caption{Core component ablation study on Arts, Games, and Instruments. The full UnpairGR model is compared with variants that remove the shared trunk (w/o ST), shared codebooks (w/o SC), or unpaired objective (w/o UP).}
  \label{fig:rq7-component-ablation}
\end{figure}

We evaluate three UnpairGR variants: \textbf{w/o ST} replaces the shared Transformer with modality-specific transformations, \textbf{w/o SC} replaces the shared residual codebooks with modality-specific codebooks, and \textbf{w/o UP} excludes unpaired supervision.
As shown in Figure~\ref{fig:rq7-component-ablation}, removing any component degrades performance across all three datasets. The largest decline occurs without the shared Transformer, reducing the macro-averaged HR@10 and NDCG@10 from $0.1477$ and $0.1040$ to $0.1170$ and $0.0813$, respectively. This pronounced degradation indicates that a shared pre-quantization transformation is critical for aligning heterogeneous observations within a coherent semantic space. Replacing the shared codebooks with modality-specific ones or unpaired supervision also causes substantial drops. These results confirm that shared representation learning, shared quantization, and unpaired supervision play complementary roles in UnpairGR.

\subsection{Efficiency and Convergence Analysis (RQ6)}
\label{sec:efficiency-convergence}

\begin{figure}[!tbp]
  \centering
  \includegraphics[width=\columnwidth]{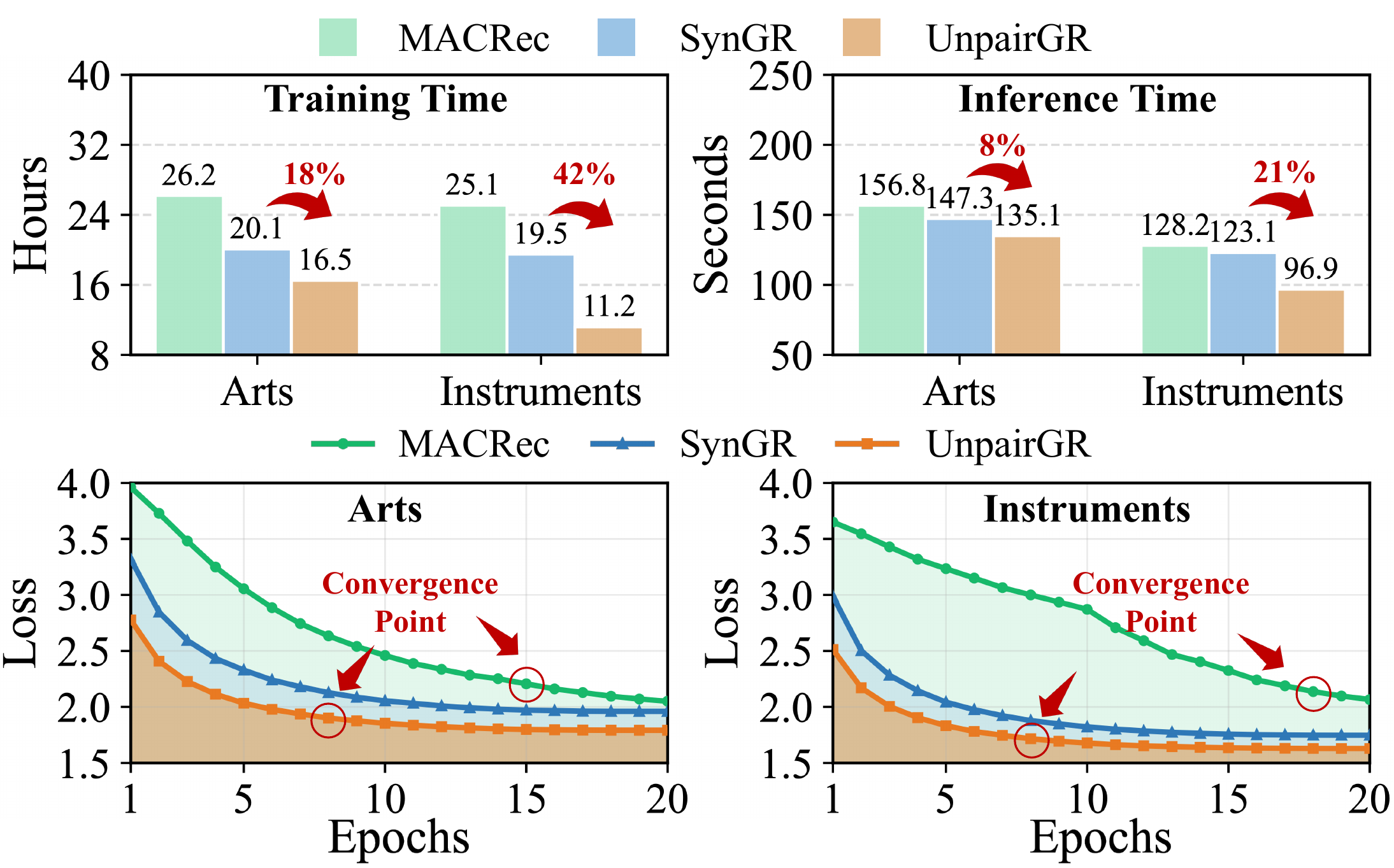}
\caption{Efficiency and convergence on Arts and Instruments. Training time, inference latency, and loss trajectories compare UnpairGR, MACRec, and SynGR under a shared protocol. Efficiency is measured using six NVIDIA RTX 4090 GPUs.}
  \label{fig:rq6-efficiency-convergence}
\end{figure}

Figure~\ref{fig:rq6-efficiency-convergence} compares UnpairGR with MACRec and SynGR under the same measurement protocol. Relative to SynGR, UnpairGR reduces training time by $18\%$ on Arts and $42\%$ on Instruments, while lowering inference latency by $8\%$ and $21\%$, respectively. The larger gains on Instruments suggest that the shared architecture scales favorably as the recommendation task becomes more demanding.
UnpairGR also converges earlier and reaches a lower final loss on both datasets, indicating stable and efficient optimization. These results show that jointly learning from paired and unimodal observations does not incur prohibitive overhead; instead, the shared Transformer and residual codebooks enable efficient training and inference while preserving optimization behavior.
\label{sec:component-analysis}

\section{Conclusion}

Missing modalities expose a weakness in multimodal generative recommendation: pre-quantization errors can alter semantic identifier sequences. UnpairGR addresses this by making the tokenizer modality-agnostic. With modality-specific projections, a shared Transformer and shared residual codebooks, and paired/unpaired training, UnpairGR learns a discrete vocabulary that remains usable under incomplete observation while preserving strong fully observed performance and practical training/inference efficiency. Component ablations show that the gains arise not from any auxiliary objective, but from the interaction of a shared continuous manifold, shared discrete vocabulary, and unpaired single-modality coverage.

\begingroup
\def\sloppy{\tolerance 3000 \hbadness 3000 \emergencystretch 2em
  \hfuzz .5pt \vfuzz .5pt \parfillskip=0pt plus 0.35\columnwidth}%
\bibliography{aaai2027}
\endgroup

\end{document}